\documentclass[prl,aps,superscriptaddress]{revtex4}
\usepackage{times} 
\usepackage{amsfonts}
\usepackage{amssymb}
\usepackage[dvips]{graphicx}

\usepackage[T1]{fontenc}
\usepackage[latin2]{inputenc}

\usepackage{amsmath}
\usepackage{bbm}

\def\>{\rangle}
\def\<{\langle}

\begin{document}


\title{Scalable Noise Estimation with Random Unitary Operators}

\author{Joseph Emerson}
\affiliation{Perimeter Institute for Theoretical Physics,\\
Waterloo, Ontario, Canada}
\author{Robert Alicki}
\affiliation{Institute of Theoretical Physics and Astrophysics,
University of Gda\'nsk, \\ Wita Stwosza 57, PL 80-952 Gda\'nsk,
Poland}
\author{Karol {\.Z}yczkowski}
\affiliation{Perimeter Institute for Theoretical Physics,\\
Waterloo, Ontario, Canada} \affiliation{Institute of Physics,
Jagiellonian University, ul. Reymonta 4,
  31-059 Cracow, Poland}
\affiliation{Center for Thoretical Physics, Polish Academy of
Sciences,
  Al Lotnik{\'o}w 32, 02-668 Warsaw, Poland}

\date{March 24, 2005}

\begin{abstract}
We describe a scalable stochastic method for the experimental
measurement of generalized fidelities characterizing the accuracy
of the implementation of a coherent quantum transformation. The
method is based on the motion reversal of random unitary
operators.  In the simplest case our method enables direct
estimation of the average gate fidelity. The more general
fidelities are characterized by a universal exponential rate of
fidelity loss. In all cases the measurable fidelity decrease is
directly related to the strength of the noise affecting the
implementation -- quantified by the trace of the superoperator
describing the non--unitary dynamics. While the scalability of our
stochastic protocol makes it most relevant in large Hilbert spaces
(when quantum process tomography is infeasible), our method should
be immediately useful for evaluating the degree of control that is
achievable in any prototype quantum processing device. By varying
over different experimental arrangements and error-correction
strategies additional information about the noise can be
determined.
\end{abstract}

\maketitle

\section{Introduction}

The characterization and elimination of decoherence and other
noise sources has emerged as one of the major challenges
confronting the coherent experimental control of increasingly
large multi-body quantum systems. Decoherence arising from
undesired interactions with background (or environment) systems
and imprecision in the classical control fields lead to severe
limits on the observation of mesoscopic and macroscopic quantum
phenomena, such as interference effects, and, in particular, the
realization of quantum communication and computation algorithms.
Measurement of the strength and other detailed properties of the
noise mechanisms affecting a physical implementation is a critical
part of optimizing, improving, and benchmarking the physical
device and experimental protocol \cite{Nicolas,Yaakov}. Moreover,
in the case of quantum devices capable of universal control,
knowledge of specific characteristics of the noise enables the
selection and optimization of passive and active error-prevention
strategies \cite{Knill,KLZ,VL,AB,Kempe,CN}.

The exact method for characterizing the noise affecting an
implementation is quantum process tomography (QPT) \cite{CN}. Let
$D$ denote the dimensionality of the Hilbert space (constituted,
e.g., from $n_q = \log_2(D)$ qubits). For QPT, the desired
transformation (usually a unitary operator) must be applied to
each member of a complete set of $D^2$ input states (spanning the
state space), followed by tomographic measurement of the output
state. This allows for a complete reconstruction of the
superoperator (completely positive linear map) representing the
imperfect implementation of the target transformation. From this
superoperator the cumulative noise superoperator can be extracted
from conventional analysis of the matrix. The QPT approach to
noise estimation suffers from several practical deficiencies.
First, often the intrinsic properties of the noise operators are
of interest, but the noise superoperator determined from QPT will
depend on the symmetries between the noise mechanisms and the
choice of target transformation. Second, the number of experiments
that must be carried out grows exponentially in the number of
qubits $D^4 = 2^{4n_q}$. Third, conventional numerical analysis of
the tomographic data requires the manipulation of matrices of
exponentially increasing dimension ($D^2 \times D^2$). For these
last two reasons QPT becomes infeasible for processes involving
more than about a dozen qubits, far fewer than the one thousand or
so qubits required for the fault-tolerant implementation of
quantum algorithms that outperform conventional computation. Hence
the infeasibility of complete noise estimation via tomography
prompts the question of whether there exist efficient methods by
which specific features of the noise may be determined.

We show below that the overall noise strength and the associated
accuracy of an implementation may be estimated by a scalable
experimental method. Specifically we show that the average gate
fidelity (\ref{avegatefid}), and some more generalized fidelities
described below, can be estimated directly with an accuracy
$O(1/\sqrt{DN})$ where $N$ is the number of independent
experiments. This method provides a solution to the important
problem of efficiently measuring which member of a set of
experimental configurations and algorithmic techniques produces
the most accurate implementation of an arbitrary target
transformation. By varying over different experimental methods and
noise-reduction algorithms and then directly measuring the
variation in the associated fidelity this method enables
estimation of more detailed characteristics of the noise.

\section{Efficient Estimation of the Average Gate Fidelity}

A convenient starting point for our analysis is the average gate
fidelity
\begin{equation} \label{avegatefid}
\overline{F_g}(\Lambda) \equiv \mathbb{E}_\psi
\left(F_{\mathrm{g}}(U, \Lambda, \psi) \right) \equiv \int d\psi
\; \< \psi |U^{-1} ( \Lambda( U |\psi \>\< \psi  | U^{-1}) ) U |
\psi \>
\end{equation}
where
\begin{equation} \label{krausform}
\Lambda(\rho) = \sum_k A_k \rho A_k^\dagger
\end{equation}
is a completely positive (CP) map characterizing the noise.
The gate fidelity $F_g$ is the inner-product of the state obtained
from the actual implementation with the state that would be
ideally obtained under the target unitary. The measure $d\psi$
denotes the natural, unitarily invariant (Fubini-Study) measure on
the set of pure states and hence the average gate fidelity
provides an indicator that is independent of the choice of initial
state. If the implementation is perfect then $\overline{F_g}=1$
and under increasing noise $\overline{F_g}$ decreases. Due to the
invariance of the Fubini-Study measure the average fidelity
depends only on the noise operator and can been expressed in the
form \cite{H3,Bowdery,Nielsen}
\begin{equation}
\label{knownaverage}
 \overline{F_{\mathrm{g}}}(\Lambda) = \frac{\sum_k |\mathrm{Tr}(A_k)|^2 + D}{D^2
 +D}.
\end{equation}
Hence the average fidelity can be determined if the noise operator
is known. The noise operator can be determined experimentally by
measuring the CP map $\Lambda(U \cdot U^{-1})$ tomographically and
then factoring out the inverse of the target map $U^{-1} \cdot U$.
This procedure has been carried out recently for 3 qubits in
recent a implementation of the quantum Fourier transform using
liquid-state NMR techniques \cite{Yaakov}. As noted above, this
method requires $\mathcal{O}(D^4)$ experiments and the
conventional manipulation of matrices of dimension $D^2 \times
D^2$. Recently Nielsen has proposed a method \cite{Nielsen} for
the direct measurement of $ \overline{F_\mathrm{g}}$ that requires
$D^4$  experiments but analysis of matrices of dimension only $D
\times D$ (rather than $D^2 \times D^2$).

We now describe how the average gate fidelity (\ref{avegatefid})
can be estimated accurately from a simple experimental protocol.
Our method requires the physical implementation of the ``motion
reversal" transformation $U^{-1} U | \psi \> \< \psi | U^{-1} U$
on an arbitrary state $| \psi \> \< \psi |$.
Under this transformation, the CP map $\Lambda$ in the gate
fidelity (\ref{avegatefid}) can be interpreted as the decoherence
and experimental control errors arising under the imperfect
implementation of the motion reversal experiment, i.e., $\Lambda =
\Lambda_{U^{-1} U }$, rather than as the noise associated with
only the forward transformation $U$, i.e, $\Lambda =
\Lambda_{U}$. 
The key idea is to choose the target transformation $U$ randomly
from the Haar measure \cite{PZK98}. This will earn us the
advantage of the concentration of measure in large Hilbert spaces,
as described further below, and leads to a universal form of the
gate fidelity depending only on the intrinsic strength of the
cumulative noise. This universal form will allow us to evaluate
the average fidelity for more generalized motion reversal
protocols.


Our starting point is the gate fidelity uniformly averaged over
all unitaries,
\begin{equation}
 \mathbb{E}_U(F_{\mathrm{g}})
 = \int_{U(D)} dU \; \mathrm{Tr}[ \rho U^{-1} \Lambda( U \rho U^{-1}) U ],
\end{equation}
where in the above $dU$ denotes the unitarily-invariant Haar
measure on $U(D)$ and $\rho = |\psi \> \< \psi |$.
In order to evaluate this integral we use the superoperator
representation of the map  (\ref{krausform}),
\begin{equation}  \label{superop}
\hat{\Lambda} = \sum_k A_k \otimes A_k^*,
\end{equation}
and similarly $\hat{U} = U \otimes U^*$, where $^*$ denotes
complex conjugation.
The Haar averaged gate fidelity takes the form
\begin{eqnarray}
  \mathbb{E}_U(F_{\mathrm{g}}) & = & \mathrm{Tr} \left( \rho
 \left[ \int dU \ \hat{U} \hat{\Lambda} \hat{U}^{-1} \right] \rho
 \right)\\
 & = & \mathrm{Tr} \left( \rho
 \hat{\Lambda}^{\mathrm{ave}} \rho \right) = F_{\mathrm{g}}(\hat{\Lambda}^{\mathrm{ave}}).
\end{eqnarray}
where $\hat{\Lambda}^{\mathrm{ave}} \equiv \int dU \ \hat{U}
\hat{\Lambda} \hat{U}^{-1}$.  As shown in the Appendix, the
Haar-averaged superoperator $\hat{\Lambda}^{\mathrm{ave}}$ is
$U(D)$-invariant and thus can be expressed as a depolarizing
channel
\begin{equation}
 \hat{\Lambda}^{\mathrm{ave}}\rho  =
p \rho + (1-p)\frac{ \mathbbm{1}}{D},
\end{equation}
(assuming $\mathrm{Tr}(\rho) = 1$) characterized by the single
``strength'' parameter
\begin{equation}
p = \frac{\sum_k |\mathrm{Tr}(A_k)|^2 - 1}{D^2 -1},
\end{equation}
where $ p \in [0,1]$ and we have made use of the fact that
$\mathrm{Tr}(\hat{\Lambda}^{\mathrm{ave}}) =
\mathrm{Tr}(\hat{\Lambda}) = \sum_k |\mathrm{Tr}(A_k)|^2$.
Direct substitution leads to
\begin{equation}
 \mathbb{E}_U(F_{\mathrm{g}})
 = F_{\mathrm{g}}(\hat{\Lambda}^{\mathrm{ave}}) = p
 +\frac{(1-p)}{D}.
 \end{equation}
Hence the gate fidelity for the Haar-averaged operator resulting
from a motion reversal experiment depends only on the single
parameter $\mathrm{Tr}(\hat{\Lambda})$ which represents the
intrinsic strength of the cumulative noise. We remark that this
result holds for general (possibly non-unital) noise. Furthermore,
suppressing the arguments of $F$ we note that the unitary
invariance of the natural measure on pure states implies the
equivalence
\begin{equation}
\overline{F_g}(\Lambda) =  \mathbb{E}_\psi( F_{\mathrm{g}}) =
\mathbb{E}_U(F_{\mathrm{g}}),
\end{equation}
and hence we recover Eq.~\ref{knownaverage}.

We now describe why and how the intrinsic noise strength
(characterized by $p$ or $\mathrm{Tr}(\hat{\Lambda})$) can be
estimated via an efficient experimental protocol. By implementing
a single target transformation $U$ that is randomly drawn from the
Haar measure, we gain the advantage of the concentration of
measure in large Hilbert spaces: the motion reversal (gate)
fidelity for the single random $U$ is exponentially close to the
Haar-averaged motion reversal (gate) fidelity. From the unitary
invariance of the Fubini-Study measure we know that
\begin{equation}
\mathbb{E}_\psi(F_{\mathrm{g}}^2)  =
\mathbb{E}_U(F_{\mathrm{g}}^2).
\end{equation}
As will be shown in Ref.~\cite{BKE}, the typical fluctuation for a
random initial state $| \psi \>$, given a fixed $U$ and $\Lambda$,
decreases exponentially with the number of qubits,
\begin{equation}
(\Delta_\psi F_g)^2 \equiv \overline{F_{\mathrm{g}}^2}  -
\overline{F_{\mathrm{g}}}^2 \leq O(1/D).
\end{equation}
Therefore it follows that,
\begin{equation}
\label{fluct} (\Delta F)_U^2 \equiv \mathbb{E}_U(F_{\mathrm{g}}^2)
- \mathbb{E}_U(F_{\mathrm{g}})^2 \leq O(1/D).
\end{equation}
Hence the fidelity under motion reversal of a single random $U$
and arbitrary (non-random) initial state is exponentially close to
the Haar-averaged fidelity
\begin{equation}
F_{\mathrm{g}}(U,\Lambda,\psi) =
F_{\mathrm{g}}(\Lambda^{\mathrm{ave}}) + O(1/\sqrt{D}) = p
+\frac{(1-p)}{D} + O(1/\sqrt{D}).
\end{equation}

The protocol is now clear: after the motion reversal sequence has
been applied experimentally, the single parameter $p$
characterizing the average gate fidelity appears as the residual
population of the initial state. Due to the invariance of the Haar
measure we may choose the initial state to be the computational
basis state $(|0\>\<0|)^{\otimes n_q}$. Hence the gate fidelity
can be determined directly from a standard readout (projective
measurement) of the final state in the computational basis. When
the noise strength is actually non-negligible (e.g., the noise
strength does not decrease as a polynomial function of $1/D$) an
accurate estimate of $p$ is possible with only a few experimental
trials. If in each of $N$ repetitions of the motion-reversal
experiment an independent random unitary is applied, then the
observed average will approach the Haar-average as
$\mathcal{O}(1/\sqrt{DN})$.


\section{Generalized Fidelities in a Discrete-Time Scenario}

More generally we imagine the ability to implement a set of
independent random unitary operators $\{ U_j\}$ and their
inverses. The entire sequence is subject to some unknown noise,
consisting of the decoherence processes and control errors
affecting the implementation. Such generalized motion reversal
sequences are relevant not only for noise-estimation, but also
have important applications in studies of fidelity decay
\cite{Emerson02} and decoherence rates \cite{ALPZ04} for quantum
chaos and many-body complex systems.

We first consider the fidelity loss arising under an iterated
motion reversal sequence of the form
\begin{equation}
\rho(n) = \hat{U}_n^{-1} \hat{\Lambda} \hat{U}_n \dots \
\hat{U}_2^{-1} \hat{\Lambda} \hat{U}_2 \ \hat{U}_1^{-1}
\hat{\Lambda} \hat{U}_1 \rho(0),
\end{equation}
where here $\hat{\Lambda}_j = \hat{\Lambda}_{U_j^{-1} U_j}$
denotes the cumulative noise from the motion reversal of $U_j$ and
we now allow arbitrary (possibly mixed) initial states $\rho(0)$.
The fidelity of this iterated transformation is,
\begin{equation}
F_n(\psi,\{U_j\}) = \mathrm{Tr}\left( \rho(0) \hat{U}_n^{-1}
\hat{\Lambda}_n \hat{U}_n \dots \hat{U}_1^{-1} \hat{\Lambda}_1
\hat{U}_1 \rho(0) \right).
\end{equation}
Averaging over the Haar measure for each $U_j$ takes the form,
\begin{eqnarray}
 \overline{F_n} \equiv \mathbbm{E}_{\{U_j\}}(F_n(\psi,\{U_j\}))
& \equiv & \int_{U(D)^{\otimes n}} \left( \Pi_{j=1}^n dU_j \right)
F_n(\psi,\{U_j\}) \\
& = & \mathrm{Tr}\left( \rho(0) \left[ \Pi_{j=1}^n
\hat{\Lambda}_j^{\mathrm{ave}} \right] \rho(0) \right) ,
\end{eqnarray}
where $dU_j$ denotes the Haar measure and we have defined the Haar
averaged noise operator,
\begin{equation}
\hat{\Lambda}_j^{\mathrm{ave}} \equiv
\mathbbm{E}_{U_j}(\hat{\Lambda}_j) \equiv \int_{U(D)} dU
\hat{U}^{-1} \hat{\Lambda}_j \hat{U}.
\end{equation}
As noted above and shown in the Appendix,
$\hat{\Lambda}^{\mathrm{ave}} \equiv \int dU \hat{U} \hat{\Lambda}
\hat{U}^{-1}$ is a depolarizing channel
\begin{equation}\label{lambdaave}
 \hat{\Lambda}_j^{\mathrm{ave}}\rho  =
p_j \rho + (1-p_j) \frac{\mathbbm{1}}{D},
\end{equation}
with strength parameter
\begin{equation}
p_j = \frac{\mathrm{Tr}(\hat{\Lambda}_j) - 1}{D^2 -1}.
\end{equation}
Because each $U_j$ is random, we can further simplify this result
by assuming that the cumulative noise for each $U_j$ has the same
strength $p_j = p$, in which case we obtain for arbitrary noise a
universal exponential decay of the averaged fidelity
\begin{equation}
 \overline{F_n} = p^n \mathrm{Tr}[\rho(0)^2] + \frac{(1-p^n)}{D}.
\end{equation}
depending only on the noise strength. In the limit of large $n$,
we see that $\overline{F_n} \rightarrow D^{-1}$, as may be
expected from the average fidelity between random states
\cite{ZS05}. Most importantly, due to the concentration of measure
($\ref{fluct}$), for large $D$ the fidelity loss under iterated
motion reversal of a single sequence of random unitary operators
will be exponentially close to the Haar-average, and hence the
noise strength can be estimated with only a few experimental runs.

Another important generalized fidelity is the one obtained under
the imperfect `Loschmidt echo' sequence \cite{Pastawski,Emerson02}
\begin{equation}
\rho(n) = \hat{U}_1^{-1} \dots \hat{U}_n^{-1} \hat{\Lambda}_n
\hat{U}_n \dots \hat{\Lambda}_1 \hat{U}_1 \rho(0),
\end{equation}
where the superoperator $\hat{\Lambda}_j$ represents the
cumulative noise during the implementation of each $U_j$. The
fidelity between the initial state and final state in the
Loschmidt echo experiment takes the form,
\begin{equation}
F_n^{\mathrm{echo}}(\psi,\{\Lambda_j\},\{U_j\}) =
\mathrm{Tr}\left( \rho(0) \hat{U}_1^{-1} \hat{U}_2^{-1} \dots
\hat{U}_n^{-1} \hat{\Lambda}_n \hat{U}_n \dots \hat{\Lambda}_1
\hat{U}_1 \rho(0) \right).
\end{equation}
Moving to the interaction picture we define
\begin{equation}
\hat{\Lambda}_j(j) =  \hat{U}_1^{-1}  \dots \hat{U}_j^{-1}
\hat{\Lambda}_j \hat{U}_j \dots  \hat{U}_1,
\end{equation}
so that,
\begin{equation}
F_n^{\mathrm{echo}}(\psi,\{\Lambda_j\},\{U_j\}) =
\mathrm{Tr}\left( \rho(0) \hat{\Lambda}_n(n)
\hat{\Lambda}_{n-1}(n-1) \dots \hat{\Lambda}_1(1) \rho(0) \right).
\end{equation}
From the invariance of the Haar measure the average fidelity
simplifies to
\begin{equation}
\overline{F_n^{\mathrm{echo}}} = \mathrm{Tr}\left( \rho(0)
\hat{\Lambda}_n^{\mathrm{ave}} \hat{\Lambda}_{n-1}^{\mathrm{ave}}
\dots \hat{\Lambda}_1^{\mathrm{ave}}  \rho(0) \right)
\end{equation}
with $\hat{\Lambda}_j^{\mathrm{ave}}$ given by
Eq.~\ref{lambdaave}. As before, we can simplify this result by
assuming that the cumulative noise for each step has the same
strength ($p_j=p$), in which case we obtain for arbitrary noise a
universal exponential form for the decay of fidelity
\begin{equation}
\label{expdecay} \overline{F_n^{\mathrm{echo}}}(p) = p^n
+\frac{(1-p^n)}{D}.
\end{equation}

A generalized version of this Loschmidt echo that is more relevant
to noise estimation is one for which noise appears in both the
forward and backward sequence of the motion reversal. The
associated fidelity is,
\begin{equation}
F_n^{\mathrm{gen}}(\psi,\Lambda,\{U_j\}) = \mathrm{Tr}\left(
\rho(0) \hat{\Lambda} \hat{U}_1^{-1} \hat{\Lambda} \hat{U}_2^{-1}
\dots \hat{\Lambda} \hat{U}_n^{-1} \hat{\Lambda} \hat{U}_n \dots
\hat{\Lambda} \hat{U}_1 \rho(0) \right).
\end{equation}
While we have not directly evaluated the average of this fidelity
analytically in the general case, for the special case of unitary
noise we have analytic and numerical evidence supporting the
relation
\begin{equation}
F^{\mathrm{gen}}_n \simeq F^{\mathrm{echo}}_{2n}
\end{equation}
for large $n$, which we conjecture should hold under general
noise.


\section{Generalized Fidelities for Continuous-Time Weak Noise}


We describe our system by the Markovian Master Equation
\cite{GKS,Al}
\begin{equation}
\frac{d}{dt}\rho = -i[H_C(t),\rho] + \epsilon{\hat L}(\rho)
\label{MME}
\end{equation}
where $H_C(t)$ governs a controlled  reversible part of the
dynamics and the generator
\begin{equation}
{\hat L} \, \rho \equiv L(\rho)= -i[H,\rho]+\frac{1}{2} \sum
_{\alpha}\bigl( [V_{\alpha} ,\rho V_{\alpha}^{\dagger}]
+[V_{\alpha}\rho , V_{\alpha}^{\dagger}]\bigr) \label{GKLS}
\end{equation}
with the condition $\mathrm{Tr}H =\mathrm{Tr}V_{\alpha}=0$ (which
fixes the decomposition of ${\hat L}$ into Hamiltonian and
dissipative parts \cite{Al}) describes all sources of
imperfections and noise. Here $0 <\epsilon \ll 1$ is a small
parameter characterizing noise strength.

The time dependent fidelity of the initial state $\phi$
is given by
\begin{equation}
F_{\phi}(t) = \<\phi| {\bf T}\exp\Bigl\{ \epsilon\int_0^t {\hat
L}(s)ds\Bigr\}(|\phi\>\<\phi|)|\phi\> \label{fid}
\end{equation}
where ${\bf T}$ denotes the chronological order, and
\begin{equation}
{\hat L}(s)= {\hat U}^{\dagger}(s,0){\hat L}{\hat U}(s,0), \quad
 U(t,s) =  {\bf T}\exp\Bigl\{ -i\int_s^t H_C(u) du\Bigr\} .
\label{gen}
\end{equation}
Using the notation
\begin{equation}
{\hat \Gamma}(t) =  {\bf T}\exp\Bigl\{ \epsilon\int_0^t {\hat
L}(s)ds \Bigr\}\label{prop}
\end{equation}
we can write down the following "cumulant expansion" of the
dynamics with respect to the small parameter $\epsilon$
\begin{equation}
{\hat \Gamma}(t)= \exp\Bigl\{ \epsilon {\hat K}_1(t) + \epsilon^2
{\hat K}_2(t) + \cdots\Bigr\}\ . \label{cum}
\end{equation}
Using the Wilcox formula for the matrix-valued functions
\begin{equation}
\frac {d}{dx}\exp A(x)= \Bigl( \int_0^1 \exp(\lambda A(x))
\frac{d}{dx} A(x)\exp(-\lambda A(x))d\lambda\Bigr) \exp A(x)
\label{Wil}
\end{equation}
one obtains
\begin{equation}
{\hat K}_1(t) = \int_0^t {\hat L}(s) ds, \quad  {\hat K}_2(t) =
\frac{1}{2}\int_0^t ds\int_0^s du [{\hat L}(s),{\hat L}(u)] \ .
\label{cum1}
\end{equation}
We assume now the following {\sl ergodic  hypothesis}: a) the
ergodic mean exists and is equal to the Haar average
\begin{equation}
\lim_{T\to\infty} \frac{1}{T}\int_0^T {\hat L}(t)dt = {\hat
L}^{\mathrm{ave}} = \int_{U(D)} dU \, \, {\hat U}{\hat L}{\hat
U}^{\dagger}\ , \label{erg}
\end{equation}
b) the fluctuations $\delta{\hat L}(t)\equiv {\hat L}(t)-{\hat
L}^{\mathrm{ave}}$ around ergodic mean are {\sl normal}, i.e. for
long $t$
\begin{equation}
\|\int_s^{s+t}\delta {\hat L}(u)du \|\sim t^{1/2} . \label{norm}
\end{equation}
These conditions are satisfied , for instance if the
time-dependent dynamics $t\mapsto U(t)$ can be modelled by a
random walk on the group $U(D)$ or by a trajectory on $U(D)$ given
by a certain deterministic dynamics with strong enough ergodic
properties. The norm of ${\hat K}_2(t)$ can be estimated using
(\ref{norm})
\begin{equation}
\|{\hat K}_2(t)\| = \frac{1}{2}\|\int_0^t ds\int_0^s du
\Bigl([\delta{\hat L}(s),\delta{\hat L}(u)] + [\delta{\hat
L}(s),{\hat L}^{\mathrm{ave}}]+[{\hat
L}^{\mathrm{ave}},\delta{\hat L}(u)]\Bigr)\|\sim t^{3/2} .
\label{K2}
\end{equation}
 Therefore
for small enough $\epsilon$ and long enough times $t$ such that
$\epsilon t$ is fixed the first term dominates and we can write
\begin{equation}
{\hat \Gamma}(t)\simeq \exp\bigl( \epsilon \int_0^t {\hat
L}(s)ds\bigr) \ . \label{cum2}
\end{equation}
Then replacing ${\hat\Gamma}(t)$ by $\exp (\epsilon {\hat
L}_{av}t)$ and using the explicit expression
(\ref{dgen},\ref{dgen1}) we obtain the universal exponential decay
of the fidelity
\begin{equation}
F_{\phi}(t) \simeq e^{-\gamma t} +\frac{1}{D} \bigl(1- e^{-\gamma
t}\bigr), \quad \gamma = \frac {D}{2(D^2 -1)}\sum_{\alpha}{\rm
 tr}(|V_{\alpha}|^2) \ .
\label{fidfin}
\end{equation}

\section{Discussion}

We have described how generalized Haar-averaged fidelities may be
directly estimated with only a few experimental measurements. By
implementing a motion reversal sequence with a Haar-random unitary
transformation, the observed fidelity decay provides a direct
experimental estimate of the intrinsic strength of the noise.
Moreover, because the target transformation is a Haar-random
unitary, the cumulative noise measured by this method will not be
biased by any special symmetries of the target transformation.

The only inefficiency of our protocol is the requirement of
experimentally implementing a Haar-random unitary: the
decomposition into elementary one and two qubit gates requires an
exponentially long gate sequence \cite{PZK98}. However, the
randomization provided by Haar-random unitary operators may be
unnecessarily strong and this leads to the open question of
whether efficient sets of random unitaries, e.g. the random
circuits studied in Refs.~\cite{Emerson03,ELL}, can provide an
adequate degree of randomization for the above protocols.
Indeed the experimental results of Ref.~\cite{Yaakov} suggest that
even a structured transformation such as the quantum Fourier
transform is sufficiently complex to approximately average the
cumulative noise to an effective depolarizing channel, and from
studies of quantum chaos it is known that efficient chaotic
quantum maps are faithful to the universal Haar-averaged fidelity
decay under imperfect motion-reversal \cite{Emerson02}. While more
conclusive evidence is needed to answer this question, it appears
likely that the inefficiency associated with implementing
Haar-random unitary unitary operators may be overcome.


An additional question is whether the implementation of random
unitary operators (e.g., Haar-random unitary operators or even
efficient random circuits) leads to an even stronger form of
averaging. We have throughout our analysis made the usual
assumption that the noise superoperator $\Lambda$ is independent
of the specific target transformation but depends only on the
duration of the experiment.
However it is known that the actual noise in general depends
sensitively on the choice of target transformation $U$. Moreover,
the cumulative noise operator generally also depends on the
particular sequence of elementary one and two qubit gates applied
to generate $U$. For example, the implementation of the quantum
Fourier transform \cite{Yaakov,Nicolas} will generate very
different cumulative noise than the trivial implementation of the
identity operator $U = {\mathbbm 1}$ for the same time $\tau$.
However, it appears likely that the cumulative noise operators,
and in particular their intrinsic noise strength, under a specific
but random gate sequence should become concentrated about an
average value depending only on the length of the sequence. If
this is the case, then the usual assumption that the noise is
independent of the actual gate sequence becomes statistically well
motivated, and the measured fidelity under motion reversal can
provide a benchmark of an intrinsic noise strength that is fully
independent of the target unitary.


{\bf Note added in proof:} additional evidence for the conjectured
relation (31) can be found in Ref.~\cite{Bettelli}.

\section{APPENDIX: Haar Averaged Superoperators}

We consider a linear superoperator ${\hat\Lambda}$ acting on the
space ${\bf M}_D$ of $D\times D$ complex matrices treated as a
Hilbert space with a scalar product $(X,Y) =
\mathrm{Tr}(X^{\dagger} Y)$. The superoperator  ${\hat\Lambda}$
has a $D^2 \times D^2$ dimensional matrix representation and
$\mathrm{Tr} {\hat\Lambda}$ denotes the usual sum over the
diagonal elements of the matrix. For clarity of notation we will
sometimes express the linear operation ${\hat\Lambda} \rho$ in the
form $\Lambda(\rho)$. By $\{|k\>\}$ we denote an orthonormal basis
in ${\bf C}^D$ while $\{E_{kl} = |k\>\<l|\}$ is a corresponding
basis in ${\bf M}_D$. The group $U(D)$ of unitary $D\times D$
matrices has its natural unitary representation on ${\bf M}_D$
defined by
\begin{equation}
U(D) \ni U\mapsto {\hat U}\ ,\quad {\hat U}X = U X U^{\dagger}\ .
\label{rep}
\end{equation}
This representation is reducible and implies the decomposition of
${\bf M}_D$ into two irreducible invariant subspaces
\begin{equation}
{\bf M}_D = {\bf M}_D^c\oplus {\bf M}_D^0\ ,\quad {\bf M}_D^0 = \{
X\in {\bf M}_D ; \mathrm{Tr}X =0\}\ ,\quad {\bf M}_D^c = \{ X = c
\, \mathbbm{1} \}, \label{rep1}
\end{equation}
where $c$ is an arbitrary complex number.

Any superoperator ${\hat\Lambda}$ possesses exactly two linear
$U(D)$ invariants, i.e. the linear functionals on superoperator
space which are invariant with respect to all transformation of
the form ${\hat\Lambda}\mapsto {\hat U}{\hat\Lambda}{\hat
U}^{\dagger}$ :
\begin{equation}
\mathrm{Tr}[\Lambda(\mathbbm{1})] = \sum_{k=1}^D \<
k|\Lambda(\mathbbm{1})| k\> \label{inv1}
\end{equation}
and
\begin{equation}
\mathrm{Tr}({\hat\Lambda}) \equiv \sum_{k,l=1}^D
(E_{kl},\Lambda(E_{kl})) = \sum_{k,l=1}^D \<k| \, \Lambda(E_{kl})
\,| l\> \label{inv2}
\end{equation}

\noindent {\bf Example} Take $\Lambda(X) = A X B $, then
$\mathrm{Tr} [\Lambda(\mathbbm{1})] = \mathrm{Tr}(AB)$
and $\mathrm{Tr}({\hat\Lambda})=\mathrm{Tr}(A) \mathrm{Tr}(B)$.

A $U(D)$-invariant operator satisfies
${\hat\Lambda}^{\mathrm{inv}}= {\hat
U}{\hat\Lambda}^{\mathrm{inv}}{\hat U}^{\dagger}$ for any $U\in
U(D)$. The following lemma completely characterizes
$U(D)$-invariant trace-preserving superoperators

\noindent
{\bf Lemma 1} Let ${\hat\Lambda}^{\mathrm{inv}}$ be a
$U(D)$-invariant trace-preserving operator. Then
\begin{equation}
{\hat\Lambda}^{\mathrm{inv}}\, X \equiv \Lambda^{\mathrm{inv}}(X)
= p \, X + (1-p)\, \mathrm{Tr}(X) \frac{\mathbbm{1}}{D} \ ,
\label{invform}
\end{equation}
where
\begin{equation}
p = \frac{\mathrm {Tr}({\hat\Lambda}^{\mathrm{inv}})- 1 }{D^2 -1}.
\label{invform1}
\end{equation}

\noindent
{\bf Proof} Schur's lemma implies the form
(\ref{invform}) for $U(D)$-invariant trace-preserving operators.
From the normalization
$\mathrm{Tr}[\Lambda^{\mathrm{inv}}(\mathbbm{1})] = D$ for the
trace, the detailed expression (\ref{invform1}) can be explicitly
calculated by comparing $U(D)$-invariants for both sides of
eq.(\ref{invform}). $\Box$

The Haar-averaged superoperator corresponding to the noise under
the imperfect motion-reversal protocol, averaged over all possible
unitary operators, is a $U(D)$-invariant superoperator
\begin{equation}
{\hat\Lambda}^{\mathrm{ave}} = \int_{U(D)} dU \  {\hat
U}{\hat\Lambda}{\hat U}^{\dagger}\ . \label{av}
\end{equation}
where $dU$ is the normalized Haar measure on $U(D)$. Using Lemma 1
we can easily compute the averaged form of the dynamical map for
both the Schr\"odinger operator
\begin{equation}
\Lambda(\rho) = \sum  _{\alpha} A_{\alpha}\rho
A_{\alpha}^{\dagger}\ ,\quad
\sum_{\alpha}A_{\alpha}^{\dagger}A_{\alpha}= \mathbbm{1}
\label{dynmap}
\end{equation}
and for the semigroup generator
\begin{equation}
{\hat L} \, \rho \equiv L(\rho) = -i[H,\rho]+\frac{1}{2} \sum
_{\alpha}\bigl( [V_{\alpha} ,\rho V_{\alpha}^{\dagger}]
+[V_{\alpha}\rho , V_{\alpha}^{\dagger}]\bigr) \label{GKLS1}
\end{equation}
with the condition $\mathrm{Tr}H =\mathrm{Tr}V_{\alpha}=0$ which
fixes the decomposition of ${\hat L}$ into Hamiltonian and
dissipative parts. From the fact that $\mathrm
{Tr}({\hat\Lambda}^{\mathrm{ave}}) = \mathrm {Tr}({\hat\Lambda})$
we obtain
\begin{equation}
\Lambda^{\mathrm{ave}} (\rho) =  p \, \rho + (1-p)
\mathrm{Tr}(\rho)\frac{\mathbbm{1}}{D} \label{dmap}
\end{equation}
where
\begin{equation}
p = \frac{\mathrm {Tr}({\hat\Lambda}) -1}{D^2-1}
 =  \frac{\sum_{\alpha}|\mathrm{Tr}(A_{\alpha})|^2 -1}{D^2-1}
\label{dmap1}.
\end{equation}
Similarly for the generator we obtain
\begin{equation}
{\hat L}^{\mathrm{ave}}\, \rho \equiv L^{\mathrm{ave}}(\rho)=
-{\gamma} \left(\rho -
\mathrm{Tr}(\rho)\frac{\mathbbm{1}}{D}\right) \label{dgen}
\end{equation}
where
\begin{equation}
\gamma = \frac {D}{2(D^2
-1)}\sum_{\alpha}\mathrm{Tr}(|V_{\alpha}|^2) \ . \label{dgen1}
\end{equation}

\section{Acknowledgements}

We would like to thank David Cory for the many discussions that
stimulated this work. R.A. would like to acknowledge the
hospitality of the Perimeter Institute for Theoretical Physics
where part of this work was completed.
We acknowledge financial support from the National Sciences and
Engineering Research Council of Canada, the Polish Ministry of
Science and Information Technology - grant PBZ-MIN-008/P03/2003,
and the EC grant RESQ IST-2001-37559.


\begin{thebibliography}{99}


\bibitem{Nicolas} N. Boulant, T. F. Havel, M. A. Pravia, and D. G. Cory, Phys. Rev. A 67, 042322
(2003).

\bibitem{Yaakov}
Y. Weinstein, T.F. Havel, J. Emerson, N. Boulant, M. Saraceno, D.
Cory, \emph{Quantum Process Tomography of the Quantum Fourier
Transform}, J. Chem. Phys. 121, 6117 (2004).


\bibitem{Knill} E.~Knill {\sl et. al.}, Introduction to Quantum Error Correction,
quant-ph/0207170 (2002).

\bibitem{KLZ} E.~Knill, R.~Laflamme and W.~\.Zurek, Resilient quantum computation, {\sl Science}, 279, 342
(1998).


\bibitem{VL} L.~Viola and S.~Lloyd, {\sl Phys. Rev. A }  {\bf 58}, 2733
(1998).

\bibitem{AB} D.~Aharonov and M.~Ben-Or, Fault-tolerant quantum computation with constant error, quant-ph/9906129
(1999).

\bibitem{Kempe} J. Kempe, D. Bacon, D.A. Lidar, and K.B. Whaley. Theory of
Decoherence-Free, Fault-Tolerant, Universal Quantum Computation.
Phys. Rev. A, 63:042307, 2001

\bibitem{CN} M.A.~Nielsen and I.L.~Chuang, {\sl Quantum Computation and Quantum Information}, Cambridge University Press
(2000)


\bibitem{H3} M.~Horodecki, P.~Horodecki and R.~Horodecki, {\sl Phys.Rev. A} {\bf 60},
1888 (1999).


\bibitem{Bowdery} M. D. Bowdrey, D. K. L. Oi, A. J. Short, K. Banaszek,
  and J. A. Jones, Phys. Lett. A 294, 258 (2002).

\bibitem{Nielsen} M.A. Nielsen, Phys. Lett. A 303 (4): 249-252
(2002).



\bibitem{PZK98} M. Po\'zniak, K. {\.Z}yczkowski, and M. Ku\'s,
 "Composed ensembles of random unitary matrices",
  {\it J.Phys.} {\bf  A 31}, 1059-1071 (1998).

\bibitem{BKE}
R. Blume-Kohout and J. Emerson, in preparation.

\bibitem{Emerson02}
 J.~Emerson, Y.S.~Weinstein, S.~Lloyd, and D.~G.~Cory,
Phys. Rev. Lett. {\bf 89}, 284102 (2002).

\bibitem{ALPZ04}
R. Alicki and A. {\L}ozi{\'n}ski, P. Pako{\'n}ski and
K.~{\.Z}yczkowski, "Quantum dynamical entropy and decoherence
rate",
 {\sl J. Phys.  A }{\bf 37},  5157-5172 (2004).


\bibitem{ZS05}
 K. {\.Z}yczkowski and H.--J. Sommers,
'Average fidelity between random quantum states' {\sl Phys. Rev.}
{\bf A 71}, 032313 (2005).

\bibitem{Pastawski} R. A. Jalabert, H. M. Pastawski, Phys. Rev. Lett. 86,
2490 (2001).


\bibitem{GKS} V.~Gorini, A.~Kossakowski and E.C.G.~Sudarshan, {\sl
J.Math.Phys. }{\bf 17} , 821 (1976); G.~Lindblad, {\sl
Commun.Math.Phys.} {\bf 48},  119 (1976).

\bibitem{Al}  R.~Alicki and K.~Lendi, {\sl  Quantum Dynamical Semigroups and Their Applications},  LNP 286, Springer,
Berlin (1987)





\bibitem{Emerson03}
 J.~Emerson, Y.S.~Weinstein, M. Saraceno, S.~Lloyd, and D.~G.~Cory,
\emph{Pseudo-Random Unitary Operators for Quantum Information
Processing}, Science 302: 2098-2100 (Dec 19 2003).



\bibitem{ELL}
J. Emerson, E. Livine, and S. Lloyd, Convergence Conditions for
Random Quantum Circuits, Phys. Rev. A 72, 060302 (2005).

\bibitem{Bettelli}
S. Bettelli, Phys. Rev. A 69. 042310 (2004).

\end{thebibliography}
\end{document}